# Electrical Tuning of Phase Change Antennas and Metasurfaces

Yifei Wang[1], Patrick Landreman[1], David Schoen[1,2], Kye Okabe[3], Ann Marshall[1], Umberto Celano[1,4], H.-S. Philip Wong[3], Junghyun Park[5], Mark L. Brongersma[1*]

[1] Geballe Laboratory for Advanced Materials, Stanford University, Stanford, California, USA

[2] Exponent Inc., Menlo Park, California, USA

[3] Department of Electrical Engineering, Stanford University, Stanford, California, USA

[4] IMEC, Leuven, Belgium

[5] Samsung Advanced Institute of Technology, Samsung Future Technology Campus Samsungro 130, Yeongtong-gu, Suwon, South Korea, 16678

[*] Correspondence and requests for materials should be addressed to Mark L. Brongersma (email: brongersma@stanford.edu).

**Abstract**

**The success of semiconductor electronics is built on the creation of compact, low-power switching elements that offer routing, logic, and memory functions. The availability of nanoscale optical switches could have a similarly transformative impact on the development of dynamic and programmable metasurfaces, optical neural networks, and quantum information processing. Phase change materials are uniquely suited to enable their creation as they offer high-speed electrical switching between amorphous and crystalline states with notably different optical properties. Their high refractive index has also been harnessed to fashion them into compact optical antennas. Here, we take the next important step by realizing electrically-switchable phase change antennas and metasurfaces that offer strong, reversible, non-volatile, multi-phase switching and spectral tuning of light scattering in the visible and near-infrared spectral ranges. Their successful implementation relies on a careful joint thermal and optical optimization of the antenna elements that comprise an Ag strip that simultaneously serves as a plasmonic resonator and a miniature heating stage.**

## Introduction

The development of metasurfaces for passive manipulation of optical wavefronts has witnessed tremendous progress over the last decade.[1–3] The research focus is now shifting to dynamic metasurfaces, which offer new applications and directions for explorative science as they are not bound by the same fundamental limits as static elements[4]. The creation of metasurface pixels that can dynamically alter their light scattering behavior not only requires the use of resonant structures to boost light-matter interaction, but also broad tunability of their resonant behavior. Effective ways to manipulate resonances apply mechanical motion[5–10], electrical gating to modify carrier concentrations[11–16] or to induce the Stark effect[17], electrochemistry[18–21], liquid crystals[22–24], and phase change materials[25–29]. Materials exhibiting structural phase transitions between crystalline and amorphous states show particular promise as they offer non-volatile switching. Metasurfaces with non-volatile pixels could afford convenient programming of desired functions and a reduced power consumption as pixels remain in a desired state after an external stimulus is removed. Reducing power consumption in dynamic metasurfaces with nanoscale pixel-arrays is shaping up as one of the key challenges moving forward as it will rival those of densely-integrated electronic circuitry.

In this work we will use germanium antimony telluride (GST), which is one of the most well-established phase change materials based on its use in optical storage media (e.g. compact disk) and non-volatile electronic memories[30]. It brings many desirable properties, including unity-order changes in the permittivity across a broad spectrum, retention times exceeding 10 years, robust switching over $10^9$ cycles, and short programming times of ~100 ns.[26,30–32] In order to appreciate the challenges in creating electrically-programmable GST antennas and metasurfaces, it is of value to study previous works on electronic and optical switching of this material. Crystallization of GST from the amorphous state can be achieved by simply heating it up to GST's crystallization temperature[35,36]. The reverse transformation back into the amorphous state is more challenging. It requires a rapid melt-quenching process with cooling rates exceeding $10^9$ K/s.[26,31,33] This was first accomplished with focused, short laser pulses as a heating source[29,37], widely used on optical discs, which recently facilitated demonstrations of a variety of optically-reconfigurable photonic elements[26,27], including plasmonic[38–41] and Mie-resonant antennas[42], waveguide-based modulators[43], and metasurfaces for beamsteering[44], lensing[29,45], perfect absorption[36], displays[46], and holographic elements[47]. The reversible electrical switching of photonic structures has proven to be more challenging. Electrically-controlled thermal cycling is well established to switch ultra-compact, 10-nm-scale GST memory cells.[33] Researchers found that this approach can produce highly-conductive, crystalline filaments in amorphous GST[30,34,35] and it has recently been successfully implemented for GST in ultrasmall plasmonic gaps to modulate waveguide transmission[52]. However, optical antennas are notably larger than memory cells (~ 100 nm) and their tuning is challenging as it requires a controlled phase change throughout their entire volume. This is hard to achieve as the comparatively low surface-to-volume ratio of these larger photonic structures tends to result in a slow cooling process that prevents an effective quench.[26] Nonetheless switching of thin layers of GST has recently been achieved through the use of micro-heater stages.[48–51] This enabled a new class of mixed-mode electronic-photonic devices capable of seamlessly interconverting electrical and optical signals in switchable thin films[51] and waveguide devices[50]. In this work, we completely reenvision the metallic heater stages used to switch GST so that they can also function as strongly scattering plasmonic antennas. To achieve robust and reversible electrical tuning of the resulting phase-change antennas requires a thoughtful co-optimization of their electrical, thermal and optical properties. In a second stage, we use the insights acquired in realizing dynamic phase-change antennas to also produce tunable metasurfaces.

**Main**

We first demonstrate the possibility to electrically tune the scattering properties of phase change antennas. Figure 1a schematically illustrates one of our antenna designs with a 60-nm-wide and 25-nm-thick GST nanobeam stacked on top of a 140-nm-wide and 36-nm-thick Ag nanostrip. The basic idea is that the Ag strip will deliver a plasmonic antenna function and the GST beam will bring dynamic switching functions. These two elements are electrically separated from each other by a 6-nm-thick layer of $Al_2O_3$ and connected to two sets of electrodes at both ends. The entire device sits on a 120-nm-thick $SiO_2$ layer on a Si substrate, whose approximately quarter-wavelength thickness was chosen to enhance the optical excitation of and collection from the antenna in the near-infrared. It was finally covered by a 20-nm-thick $Al_2O_3$ capping layer to protect it from oxidation. This design offers several desirable optical, electrical, and thermal properties that facilitate effective tuning of the light scattering. To perform accurate optical simulations, the optical properties of our GST material are first measured using ellipsometry (Supplementary Note 1). Our simulations show that under dark-field illumination with TM-polarized light (H field directed along the length of the nanobeam) at 800 nm, surface plasmon polaritons (SPPs) are excited at the interface between the GST and Ag (Fig. 1b). The fields are largely confined to the volume of the GST beam. This is consistent with a physical picture where excited SPPs resonate by oscillating back and forth inside the GST beam, reflecting from its side walls.[53,54] For this reason, the width of the GST beam critically controls the resonance wavelength and narrower (wider) beams can be chosen to shift the resonance to the blue (red). The good modal overlap of the SPP mode with the switching medium ensures effective tuning of the resonant scattering properties upon a phase change.

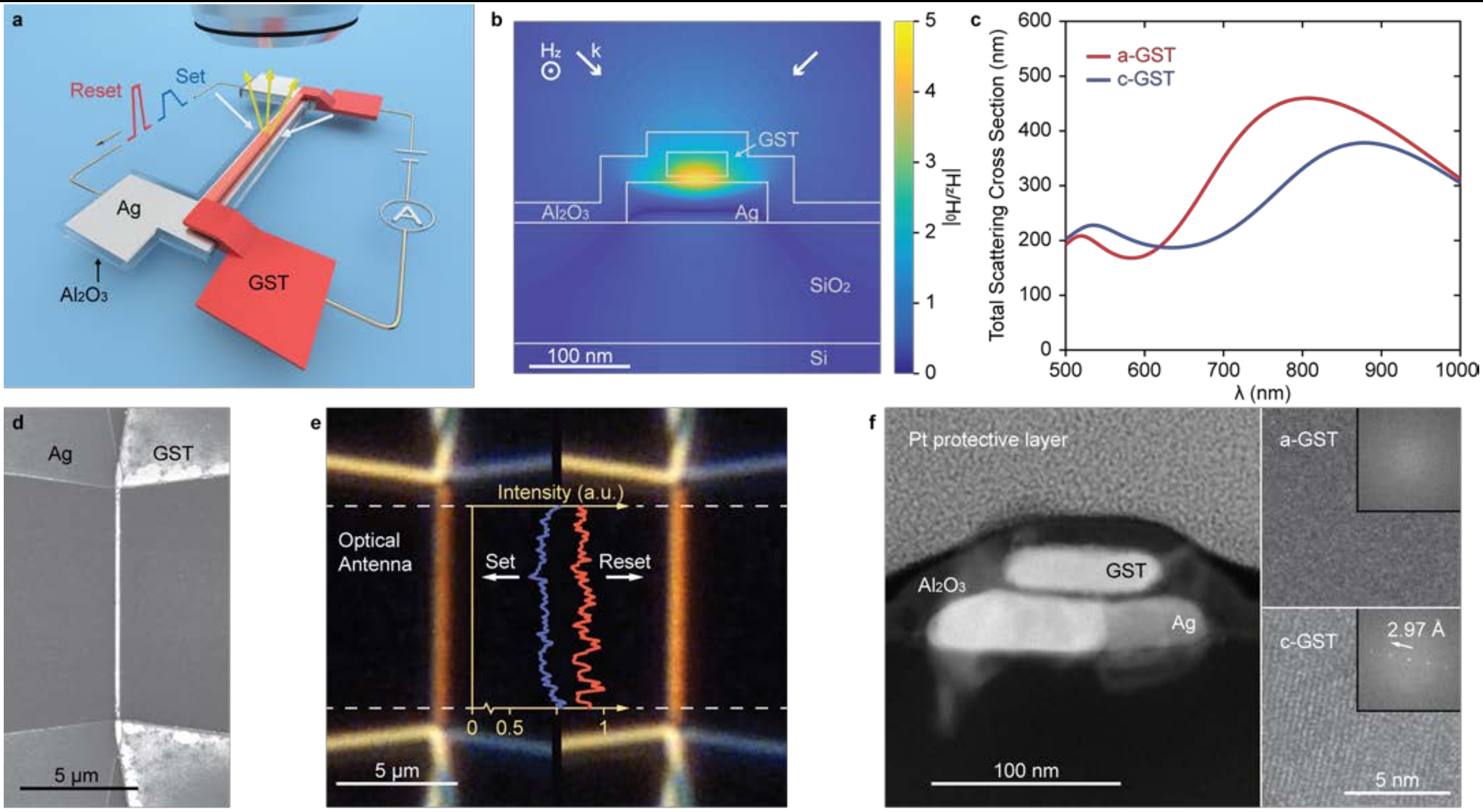

**Figure 1: Operation and optical response of an electrically-tunable phase change antenna. a.** A schematic showing an optical antenna comprised of a GST nanobeam stacked on top of an Ag nanostrip, separated by a thin $Al_2O_3$ layer. Set (blue) and Reset (red) current pulses are sent through the Ag nanostrip heater/antenna to transform a-GST to c-GST and back. The resistance of the GST beam and the scattered light from the device in dark-field, white-light illumination can be monitored to study the impact of the phase changes on the electrical and optical properties. **b.** Finite-difference time-domain simulation of the antenna with a-GST illuminated at grazing-incidence and with TM-polarized light at a wavelength of 800 nm. **c.** The simulated total scattering cross section of the optical antenna with a-GST and c-GST. **d.** Scanning electron microscopy (SEM) image of a fabricated device **e.** Dark-field, white-light scattering images of the antenna taken after a set pulse and a reset pulse, showing a notable change in scattering intensity. The scattered light intensity measured along the antenna length is plotted in the inset. **f.** Cross section TEM images. The left image shows a scanning TEM image of the stacked GST and Ag beams and the thin $Al_2O_3$ spacer. On the right are high-resolution TEM images of this type of device after applying reset and set pulses, respectively. Insets show the corresponding fast Fourier transforms (FFT) with a logarithmic intensity scale. The top cross section shows no crystallinity, while the bottom image shows the lattice fringes of c-GST.

Figure 1c shows the resonant scattering behavior for the cases where the GST is in the amorphous (a-GST) and crystalline phase (c-GST). The long wavelength resonance corresponds to the fundamental dipole resonance with one antinode in the magnetic field above the Ag strip and the short wavelength resonance corresponds to a resonance with three antinodes (Fig. S3). Both resonances exhibit clear redshifts as the phase is transformed from amorphous to crystalline due to an increase of the imaginary part of GST permittivity. In this work we will mainly focus on the stronger, fundamental resonance. The scattering cross section reduces upon switching from a-GST to c-GST due to the higher materials loss in the crystalline phase at the longer wavelength.

The changes in the scattering cross section can be directly observed in dark-field microscopy under white-light illumination. Figure 1d shows a scanning electron microscopy (SEM) image of the antenna and pads that electrically connect the GST beam and Ag strip. The optical images taken from the antenna before and after amorphization show clear differences in the scattered light intensity (Fig.1e). Both antennas show a more-or-less constant scattering intensity along the length of the beams, as expected for well-fabricated antennas.

A structural analysis of the antennas before and after switching further confirms our ability to reversibly switch the GST material between crystalline and amorphous phases. Using a focused ion beam (FIB) lift-out technique, we prepared antennas for electron microscopy analysis. Transmission electron microscopy (TEM) cross-sectional imaging (see Fig. 1f) shows the structure was successfully fabricated with the intended geometry and dimensions. High resolution transmission electron microscopy (HR-TEM) imaging of the GST nanobeams after a reset pulse exhibits no lattice fringes, consistent with an amorphous phase, while GST after a set pulse exhibits lattice fringes with a spacing that is consistent with c-GST. More details on the sample preparation and measurement analysis are discussed in Supplementary Note 4.

The Ag strip also performs important electrical and thermal functions. By running current through the beam, it can serve as a miniature hot-plate capable of heating the GST beam that sits on top. Two kinds of current pulses with different temporal profiles are sent through the Ag strips. Set pulses of a 1 μs duration and with a 20 μs trailing time are used to heat the GST above the crystallization temperature ($T_x$) and then maintain the high temperature long enough to facilitate complete crystallization.[33,55] Reset pulses feature a higher current for a shorter duration of 500 ns to rapidly raise the local temperature above the melting temperature of GST ($T_{melt}$ ~600 °C). The pulse is then switched off quickly with a 20 ns trailing time to achieve an effective melt-quench that freezes in the amorphous phase. The key challenge towards realizing reversible switching is achieving a sufficiently fast cooling rate to prevent recrystallization as the GST cools. From the literature it is established that for our thermal cycle typical crystallization times are around 80 ns.[56] We find that it is beneficial to have an Ag strip that is wider than the GST beam to ensure excellent heat dissipation to the substrate. A thermal simulation (Methods) of the device shows that a reset pulse with a 100 ns rise time, can effectively heat the GST beam to above $T_{melt}$ in about 200 ns (Fig. 2a). At the end of the pulse, the current drops with a 20 ns trailing time and this allows the GST to cool down to a temperature below 120 °C (below $T_x \approx 160$ °C) in less than 50 ns (Fig. 2b). Our 2D simulations assume that the antenna is infinitely long, which is justified for our 10 µm-long and 140-nm-wide antennas (Fig. 1d).

We further analyze the GST switching behavior by monitoring the resistance of the GST nanobeam after sending every reset and set pulse (see Fig. 2c). The resistance after each reset pulse is seen to consistently increase by at least an order of magnitude and the opposite occurs after each set pulse. This confirms that we can achieve robust, reversible switching between the a-GST and c-GST phases. The slight variations in the resistance may indicate not fully completed transformations. Additionally, we performed conductive atomic force microscopy (C-AFM) to demonstrate that the phase transition happens along the entire length of the GST nanobeam (Supplemental Note 5).

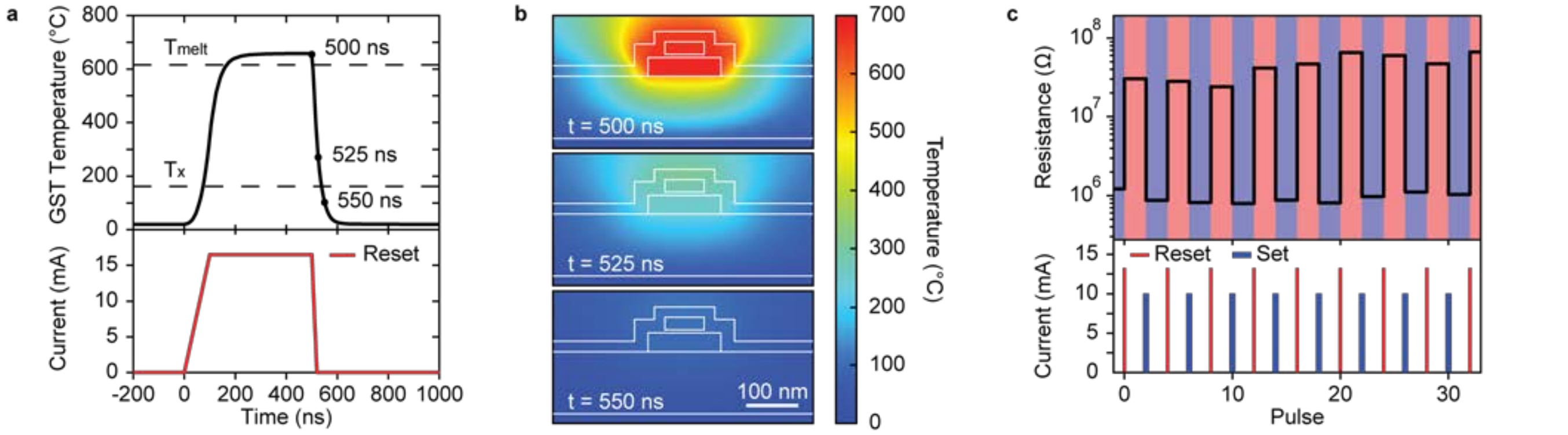

**Figure 2: Electrothermal switching of optical antennas. a.** A thermal simulation of the phase change antenna when applying a reset pulse. The temporal evolution of the average temperature of the GST nanobeam (top) is plotted along with the electrical pulse current (bottom). The melting temperature ($T_{melt}$) and crystallization temperature ($T_x$) of GST are plotted with dashed lines respectively. **b.** Cross sectional plot of the antenna temperature profile during the cooling process, the time stamps correspond to the three dots on the temperature curve in panel a. **c.** Electrical testing of an antenna. The resistance of the GST nanobeam is measured after each pulse, where the red regions show the resistance after reset pulses and blue regions indicate the resistance after set pulses. The peak pulse current is plotted at the bottom. Robust and reversible switching between crystalline and amorphous phases is confirmed by the resistance changes in excess of one order of magnitude.

Scattering spectra from the antenna are taken after each current pulse (Fig. 3a). The measured spectra show broad resonances that are peaked just above 700 nm. They qualitatively match the simulated spectra of the total scattering cross section in Fig. 1c. A notable 30% modulation in intensity is observed on resonance. The spectra for the amorphous and crystalline phase show good consistency over many cycles. The simulated spectra in Fig. 3b show good agreement when we take into account that in our experiments we only collect back-scattered light through a 50× objective with a numerical aperture of 0.6. The spectra that are simulated assuming complete amorphization and crystallization indicate a possible signal modulation exceeding 50%. An antenna with different volume fractions of amorphous material are also simulated, assuming the top part of the antenna is in the crystalline phase. We find that the spectra continuously morph from the crystalline spectrum to the amorphous spectrum as the volume fraction is increased. The measured spectra after the reset pulses in Fig. 3a agree best with simulations assuming a high (~ 80%) volume fraction of a-GST. It could be that the GST antenna does not return to the completely amorphized state after reset pulses and part of the GST material may have recrystallized during the quenching process. We confirmed that GST can indeed be switched completely to the crystalline state by sending set pulses with higher current. In that case, no further decrease of the back scattered signal is observed after multiple set pulses.

We also use a CCD camera to image the antenna in real time as we are sending switching pulses (See Fig. 1e). We integrate the intensity over the antenna area, to plot the scattered intensity against time as different set and reset pulses are delivered to the antenna (Fig. 3c). The collected signal shows signal changes of 30% over more than 100 cycles, as shown in Fig. S10.

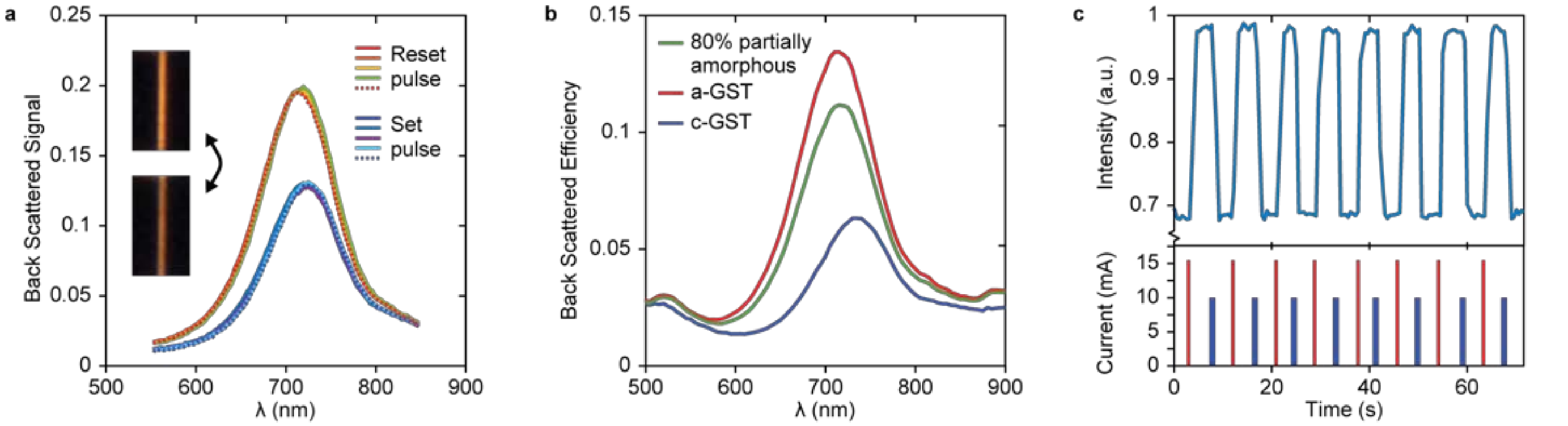

**Figure 3: In situ optical measurement of an antenna. a.** Measured spectra of the scattered light from an antenna after reset and set pulses, under dark-field illumination. The dashed lines represent spectra collected after 10 cycles of operation. The inset shows two images of the antenna under dark-field illumination after a reset pulse (top) and a set pulse (bottom). **b.** Simulated spectra of the scattered light from the antenna, collected with an objective (NA: 0.6), when the GST nanobeam is in the amorphous, crystalline and partially amorphous phase (80% is amorphous with a top 20% layer in crystalline state). **c.** Demonstration of binary operation of the antenna, with a change of 30% in scattered light intensity after reset (red) and set pulses (blue).

While the reset pulses quench the GST quickly in the above experiments, we can increase the trailing time to induce slower cooling and allow the GST to fully crystallize. By varying the temporal shape of the pulses, we can control the degree of crystallization and achieve multi-level operation. (Fig. S11)

After we demonstrated the repeatable switching of a single antenna, we move on to explore whether more complex metasurfaces can be created comprised of multiple antennas. Figure 4a shows a schematic of a metasurface designed to operate as a perfect absorber[13,36] in the crystalline phase. Here, 85-nm-wide and 35-nm-thick GST nanobeams are placed in an array with a period of 200 nm on top of a 60-nm-thick Ag strip. The Ag strip is chosen to be thick enough to prevent any light transmission.

Again a thin (6 nm) layer of $Al_3O_2$ is used as a spacer and a 50-nm-thick $Al_3O_2$ capping layer was deposited to protect the device. In Fig. 4b, we show a FDTD simulation of the metasurface with c-GST beams, under TM-polarized illumination at normal incidence. We first show the field distribution at the wavelength of 730 nm, where SPPs are excited on the Ag surface and inside the GST beams. The high fields result in significant dissipation inside the beams causing almost perfect absorption. This is confirmed by the overlaid flowlines of the Poynting vector that show efficient funneling of an incident plane wave into the lossy GST beams. The ability to achieve perfect absorption can be understood by comparing the scattered fields from a sample with just the Ag strip and one with the GST beams on top of the strip. They are seen to be perfectly out of phase and of equal amplitude. This indicates that the non-resonant path (reflection from the Ag strip) and the resonant path (storing energy in the resonant antenna modes) can destructively interfere and prevent an outgoing/reflected wave. After switching to a-GST, the metasurface reflects 30% of the light at 730 nm and the absolute modulation of the reflectance increases at longer wavelengths (Fig. 4c).

To verify the simulations, we measure reflection spectra from the metasurface at normal incidence. When the beams are in the crystalline phase the lowest reflectance of 4.3% is achieved at the wavelength of 700 nm, consistent with the simulations. After a reset pulse, the reflected light intensity at this wavelength increases to 14.5%, a more than 3-fold change. The non-zero reflectance for c-GST is attributed to the small variations in the nanobeam dimensions that disturb the delicate, destructive interference condition. The highest modulation in the reflectance of a factor 4.5 is achieved at a wavelength of 755 nm. We also collect the reflected light with a CCD camera after passing through a band-pass filter centered at 690 nm, with a full width half maximum (FWHM) of 10 nm. In Fig. 4e, the reflected signal is modulated up and down with reset and set pulses respectively, showing a good consistency with repeated cycling.

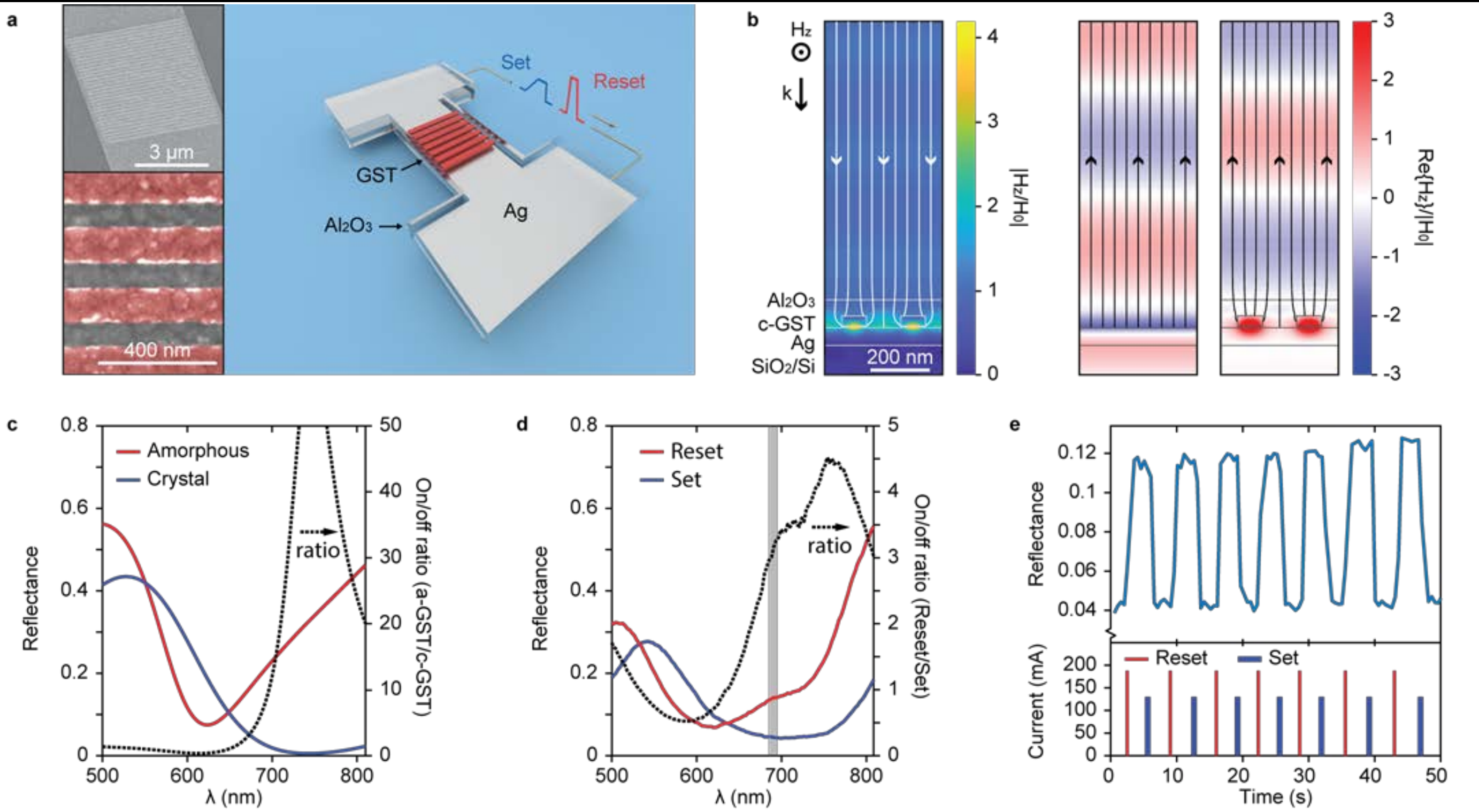

**Figure 4: An electrically-tunable phase change metasurface. a.** A schematic of the proposed metasurface is shown on the right, with a dense GST nanobeam-array placed on an optically-thick Ag strip. Reset and set pulses are sent through the Ag strip to heat the metasurface and reversibly transform it between a-GST and c-GST phases. The left top inset is an SEM image of a device. The left bottom inset is a false-colored SEM image of the GST array, where the red regions highlight the GST nanobeams. **b.** FDTD simulation of the metasurface with c-GST at normal-incidence and with TM-polarized light at a wavelength of 730 nm. The first figure plots the total field. The second and third figure shows the scattered field for the Ag strip by itself and for the case the GST nanobeams are added, displaying a π phase difference. **c.** Simulated spectra of the reflectance and transmittance of the metasurface with a-GST and c-GST beams. The ratio of the reflectance for the a-GST and c-GST states is shown by the dotted black curve. **d.** Measured spectra of the reflectance of the metasurface taken after a reset and set pulse. The on/off ratio of the reflectance (reset/set) is also shown as the black curve, with a maximum ratio of 4.5 at the wavelength of 755 nm. The FWHM of the band-pass filter used to obtain **Fig. 4e** is shown in gray. **e.** Demonstration of binary operation of a "perfect absorber" metasurface, showing a modulation of reflectance by more than 2.5-fold after reset and set pulses.

## Conclusion

In conclusion, we have demonstrated a method for realizing electrical, non-volatile, and reversible switching of phase change antennas and metasurfaces devices based on a chalcogenide phase change material. A design was created in which Ag nanostrips can conveniently function as a nano-heater and enable plasmonic antenna functions. We show an optical antenna that offers strong scattering in the visible and near-infrared spectral ranges and significant (30%) electrical modulation of the scattered light intensity. Utilizing destructive interference between two pathways, a metasurface is also demonstrated, with an electrical modulation of the reflectance by more than 4-fold. This work opens the opportunity to create a wide range of dynamic random access metasurfaces capable of programmable and active wavefront manipulation.

**Methods:**

Devices fabrication

The optical antennas and metasurface devices are fabricated following many of the same, initial fabrication steps. First, a 120-nm-thick layer of $SiO_2$ is grown by thermally oxidizing a Si substrate. Alignment markers and contact pads are subsequently defined on the substrates by photolithography. We then use electron-beam evaporation to first deposit a 5-nm-thick Ti adhesion layer, followed by a 45-nm-thick Au later and a lift-off in acetone.

The next steps to fabricate the optical antennas involve an electron-beam (e-beam) lithography to define the Ag strips, followed by e-beam evaporation of a 4-nm-thick Ti adhesion layer and a 36-nm-thick Ag layer and ultimately a lift-off. We then deposit an $Al_2O_3$ layer with a thickness of 6 nm by atomic layer deposition (ALD). Another set of contact pad are defined by photolithography, designed to connect the GST beams. A 5-nm-thick Ti adhesion layer and a 45-nm-thick Au layer are then deposited by e-beam evaporation. For the next layer, we use e-beam lithography to define the pattern, then use DC magnetron sputtering to deposit the GST from a $Ge_3Sb_2Te_6$ (GST-326) target, to achieve 24-nm-thick GST nanobeams. As the last step, a protective $Al_2O_3$ capping layer with a thickness around 20 nm is deposited by ALD (thinner on top of GST at 13 nm).

The next steps to fabricate active metasurfaces involve e-beam lithography to define the Ag strips and a deposition of a 6-nm-thick Ti adhesion layer and 54-nm-thick Ag layer. We then deposit a 6-nm-thick $Al_2O_3$ layer by ALD. E-beam lithography is used to define the GST patterns, followed by DC magnetron sputtering of GST-326, producing 34-nm-thick GST nanobeam-arrays. An $Al_2O_3$ capping layer with a thickness around 50 nm is deposited by ALD.

Finite-difference time-domain (FDTD) simulations

Optical simulations are carried out with FDTD method (FDTD Simulations, Lumerical). Two-dimensional (2D) simulations of the single optical antennas are performed with perfectly matched layer (PML) boundary condition. While a single-wavelength TFSF source is used to study the impact of an inclined illumination at 46°, we sweep the wavelength with a fixed angle of incidence to obtain the spectrum across a broad wavelength range. Monitors are placed around the device, outside of the TFSF source region to calculate the far-field profile of electromagnetic field at each wavelength. We then integrate the field intensity within a solid angle, corresponding to a numerical aperture of 0.6. To further simulate the experimental condition, we measure the collection efficiency of the objective used in experiments (Nikon TU Plan 50×/0.60) at each wavelength. This allows us to correct for the instrument response. 2D simulation of the metasurface are performed using periodic boundary conditions.

Finite element method (FEM) simulation

We also performed 2D thermal simulations a finite-element method (COMSOL Multiphysics). They are implemented for a 10 μm×10 μm region, with the boundary set at room temperature. The thermal model is discussed in more details in Supplementary Note 2.

Electrical characterization

A semiconductor device parameter analyzer is used for electrical characterization, including a pulse generator which sends electrical pulses through Ag strips to induce phase transition and measurement units to monitor the resistance of both Ag and GST strips. Please refer to Supplementary Note 3.

In-situ optical measurements

We measure the dark-field scattering spectra of the antennas and the bright-field reflection of the metasurfaces with a Nikon C2 confocal microscope. A halogen lamp is used as the illumination source. After being collected by a 50× objective (Nikon TU Plan 50×/0.60), the scattered light from devices is polarized in the TM direction with respect to the devices and is subsequently sent into either of two paths. In the first path, a Nikon DS-Fi1 camera is used to capture images and videos. Alternatively in a second path, a confocal scanner spatially selects the scattered light from the devices and sends it to a Princeton Instruments SpectraPro 2300i spectrometer (150 lines/mm, blazed for $\lambda = 500$ nm) and a Pixis Si CCD camera (−70 °C detector temperature) to analyze the spectrum. The antennas are measured under dark-field illumination at a 46° incident angle. The scattered light is collected through a 30-μm-diameter pinhole in the confocal scanner, before analysis by the spectrometer. The measured spectra are normalized to the scattered spectrum of a Lambertian reflector (Labsphere Spectralon 99%) measured under the same condition. The metasurface devices are measured under normal-incidence illumination, with an aperture stop closed to ensure a minimal beam divergence. The measured spectra are normalized to the reflection spectrum of an Ag mirror (Thorlabs PF10-03-P01). While under the microscope for in situ optical measurement, the devices are connected to a semiconductor analyzer by probe tips to perform electrical switching.

Transmission electron microscopy and sample preparation

Transmission electron microscopy was used to analyze the structural phase and physical dimensions of the devices. Details can be found in Supplementary Note 4.

Conductive atomic force microscopy

Please see the discussion in Supplementary Note 5.

## Acknowledgment

The work was supported by was supported by an individual investigator project funded by Samsung Advanced Institute of Technology. We would like to also acknowledge funding from AFOSR MURI grant (FA9550-17-1-0002). Part of this work was performed at the Stanford Nano Shared Facilities (SNSF), supported by the National Science Foundation under award ECCS-1542152. U.C. was supported in part by Fonds voor Wetenschappelijk Onderzoek—Vlaanderen (FWO).

## Author contributions

Y.W., P.L. and M.L.B. conceived the ideas for this research project. Y.W. fabricated the devices, performed optical simulations, and implemented optical microscopy measurements. D.S. and A.M. prepared the sample and carried out TEM measurements. Y.W. conducted thermal simulations and electrical measurements with the help of K.O. and H.-S.P.W. C-AFM measurements were performed by Y.W. and U.C. All authors contributed to the writing of the manuscript.

## Competing financial interests

The authors declare no competing financial interests.